\documentclass[final,5p,times,twocolumn]{elsarticle}

\usepackage{amssymb}
\usepackage{array}
\usepackage{amsthm}
\usepackage{braket}
\usepackage{array}
\usepackage{siunitx}
\usepackage{tabularx}
\usepackage{hyperref}
\usepackage{lineno}

\journal{Nuclear Instruments and Methods in Physics Research Section A}
\begin{document}
\begin{frontmatter}

\title{New application of superconductors: high sensitivity cryogenic light detectors}


\author[Sapienza,Princeton]{L.Cardani}
\ead{laura.cardani@roma1.infn.it}
\author[Sapienza,InfnRoma]{F.~Bellini}
\author[Sapienza,InfnRoma]{N.~Casali}
\author[CNR]{M.~G.~Casellano}
\author[Sapienza]{I.~Colantoni}
\author[Sapienza]{A.~Coppolecchia}
\author[Sapienza,InfnRoma]{C.~Cosmelli}
\author[Sapienza,InfnRoma]{A.~Cruciani}
\author[LNGS]{A.~D'Addabbo}
\author[INFNGe,Genova]{S.~Di~Domizio}
\author[Sapienza,InfnRoma,Zaragoza]{M.~Martinez}
\author[InfnRoma]{C.~Tomei}
\author[Sapienza,InfnRoma]{M.~Vignati}

\address[Sapienza] {Dipartimento di Fisica - Sapienza Universit\`{a} di Roma, Piazzale Aldo Moro 2, 00185, Roma - Italy}
\address[Princeton] {Physics Department - Princeton University, Washington Road, 08544, Princeton - NJ, USA}
\address[InfnRoma] {INFN - Sezione di Roma, Piazzale Aldo Moro 2, 00185, Roma - Italy}
\address[CNR] {Istituto di Fotonica e Nanotecnologie - CNR, Via Cineto Romano 42, 00156, Roma - Italy}
\address[LNGS] {INFN - Laboratori Nazionali del Gran Sasso, Assergi (L'Aquila) 67010 - Italy}
\address[INFNGe] {INFN - Sezione di Genova, Via Dodecaneso 33, 16146, Genova - Italy}
\address[Genova] {Dipartimento di Fisica - Universit\`{a} degli Studi di Genova, Via Dodecaneso 33, 16146, Genova - Italy}
\address[Zaragoza]{Laboratorio de Fisica Nuclear y Astroparticulas, Universidad de Zaragoza, Zaragoza 50009 - Spain}

\begin{abstract}
In this paper we describe the current status of the CALDER project, which is developing ultra-sensitive light detectors based on superconductors for cryogenic applications. 
When we apply an AC current to a superconductor, the Cooper pairs
oscillate and acquire kinetic inductance, that can be measured by inserting
the superconductor in a LC circuit with high merit factor. Interactions in
the superconductor can break the Cooper pairs, causing sizable variations
in the kinetic inductance and, thus, in the response of the LC circuit. The
continuous monitoring of the amplitude and frequency modulation allows
to reconstruct the incident energy with excellent sensitivity. This concept is
at the basis of Kinetic Inductance Detectors (KIDs), that are characterized
by natural aptitude to multiplexed read-out (several sensors can be tuned to
different resonant frequencies and coupled to the same line), resolution of
few eV, stable behavior over a wide temperature range, and ease in
fabrication. We present the results obtained by the CALDER collaboration
with 2$\times$2\,cm$^2$ substrates sampled by 1 or 4 Aluminum KIDs. We show that
the performances of the first prototypes are already competitive with those
of other commonly used light detectors, and we discuss the strategies for a
further improvement.
\end{abstract}


\end{frontmatter}

\section{Why new cryogenic light detectors?}
\label{sec:intro}
Background reduction is becoming more and more important for all the experiments looking for rare events, such as dark matter interactions, neutrino-less double beta decay (0$\nu$DBD), or rare $\alpha$ decays. 
Among the several technological approaches proposed for these searches, cryogenic calorimeters (or bolometers) stand out for their excellent energy resolution (of the order of 0.1$\%$) and for the possibility of probing different compounds; the crystals operated as bolometers can be grown starting from most of the possible 0$\nu$DBD candidates or rare-$\alpha$ emitters, and from several interesting targets for dark matter search. 

Bolometers can be equally sensitive to $\alpha$'s, electrons and, if the threshold is low enough, also to nuclear recoils. 
This could be an advantage, as experiments designed for a specific purpose could be competitive also in other physics sectors. 
On the other hand, the lack of tools to identify the nature of the interacting particles prevents an efficient background suppression. A possible solution consists in equipping the bolometer with a light detector, that enables particle identification exploiting the different light emission of $\alpha$'s, electrons and nuclear recoils. 
This technique has been successfully exploited for bolometers emitting scintillation light, like ZnSe~\cite{Beeman:2013vda}, ZnMoO$_4$~\cite{Beeman:2012ci}, LiMoO$_4$~\cite{Cardani:2013dia} and many others.
The scintillation crystals were coupled to light detectors made by thin Germanium disks equipped with  Neutron Transmutation Doped Germanium sensors, that show a typical intrinsic energy resolution of about 80\,eV RMS~\cite{Beeman:2013zva}.
This sensitivity is high enough to enable an efficient particle identification with scintillation bolometers, but may be not sufficient for next generation experiments.
A possible application of new light detectors with these characteristics could be the upgrade of the CUORE experiment. The TeO$_2$ bolometers used by this experiment do not scintillate. On the other hand, coupling a sensitive light detector to a TeO$_2$ would allow to measure the tiny amount of Cherenkov light emitted by electrons (about 100\,eV at 0$\nu$DBD energy) and not by $\alpha$'s, enabling the rejection of the dominant $\alpha$ background~\cite{Bellini:2014yoa,Casali:2014vvt}.

The CUPID interest group~\cite{Wang:2015taa,Wang:2015raa}, that is defining the road-map for a 0$\nu$DBD next-generation project based on bolometers, identified the features of the desired light detector: a noise RMS resolution lower than 20\,eV, high radio-purity, large active area (5$\times$5\,cm$^2$) and ease in fabricating/operating up to 1000 channels.

The growing interest in developing sensitive light detectors (that would be important also for other applications~\cite{Beeman:2013vda}) gave birth to several R$\&$D activities exploiting different technologies~\cite{Schaffner:2014caa,Willers:2014eoa,Biassoni:2015eij}. None of these technological approaches is (yet) able to fulfill all the requirements for next generation projects. For this reason, the CALDER (Cryogenic wide-Area Light Detectors with Excellent Resolution~\cite{CalderWhitePaper}) project proposes a new technique, based on Kinetic Inductance Detectors.

\section{Phonon-mediated Kinetic Inductance Detectors}
\label{sec:KIDs}
Kinetic Inductance Detectors (KIDs) are superconductors biased with an AC current~\cite{Day:2003fk}.
The energy of the electric field applied to the superconductor is stored in the kinetic energy of Cooper pairs. The oscillation of the field and, as a consequence, of the Cooper pairs, produces an additional inductance in the superconductor, called kinetic inductance L$_K$.
By coupling the superconductor with a capacitor, we can realize a RLC circuit, that acts like a resonator with resonant frequency $f_0=\frac{1}{2\pi\sqrt{LC}}$. The properties of superconductors allow to fabricate resonators with typical quality factor (Q) of the order of 10$^4$-10$^5$.

A photon interacting in the superconductor breaks Cooper pairs, increasing the average momentum per Cooper pair and, as a consequence, increasing L$_K$. 
Moreover, the breaking of Cooper pairs into quasiparticles increases the dissipation of the resonator, thus reducing Q.
Monitoring the changes in the resonance parameters (frequency and phase) we can infer the energy that was deposited by the photon interaction. This is the working principle at the basis of KIDs that, besides an excellent energy resolution ($\sim$eV) bring the advantage of being naturally multiplexed in the frequency domain; this means that hundreds of KIDs can be coupled to a single read-out wire, decreasing the number of electronics channels, as well as the heat-load on the cryogenic system.

The main limit of this technology is that KIDs can feature a maximum active area of a few mm$^2$, while next generation experiments demand for light detectors with surface of tens of cm$^2$. 
In principle, we could realize a large area light detector with an array of hundreds of KIDs. This is not a realistic option, because photons do not interact efficiently in the superconductor and, in addition, the operation of about 1000 light detectors would require a very large number of KIDs (hundreds of thousands).

For this reason, we deposit KIDs on an insulating substrate that is used to absorb the interacting photon. The absorption of a photon produces phonons, that travel in the substrate until they are absorbed by the KIDs or lost in the substrate supports or defects. 
The goal of the CALDER project is proving that this approach, proposed by Swenson et al.~\cite{swenson} and Moore et al.~\cite{moore2} for other applications, allows to realize a light detector with all the characteristics required by next-generation experiments. 

The project foresees three main phases. 
The fist stage is devoted to the development of all the necessary acquisition and analysis tools and to the optimization of the detector geometry. For this phase we decided to work with a well known material for KIDs applications, Aluminum, that will allow to reach an RMS energy resolution of about 80\,eV.
In the second stage we will move to more sensitive superconductors, such as TiN, Ti+TiN, or TiAl, in order to lower the energy resolution below 20\,eV. 
Finally, we will couple the optimized light detectors to an array of TeO$_2$ bolometer to prove the potential of this technology.

\section{Results of the first phase: Aluminum KIDs}
\label{sec:first phase}
\begin{figure}[thbp]
\centering
 \includegraphics[width=.45\textwidth, natwidth=1105, natheight=829]{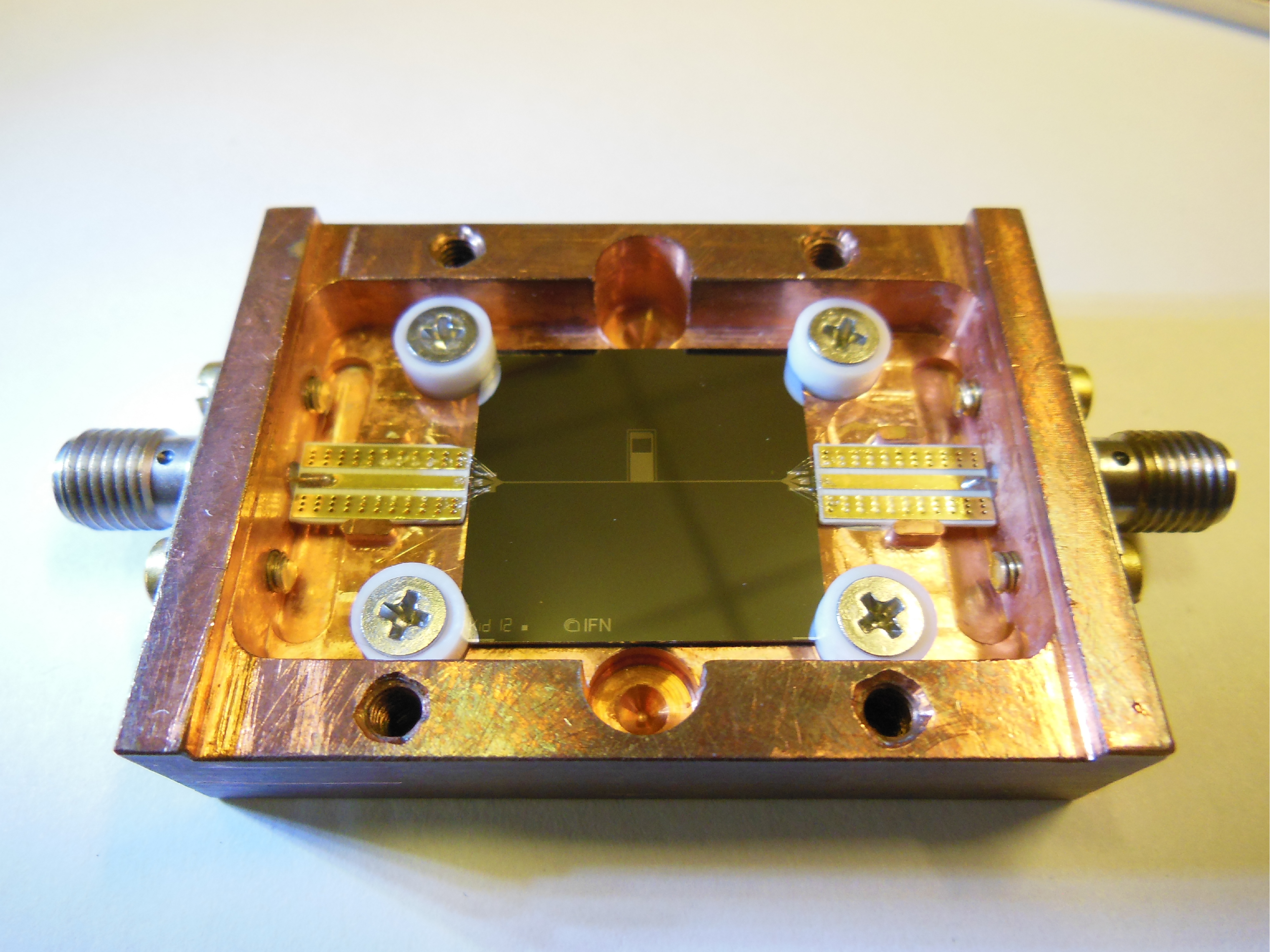}
 \caption{\label{fig:photo}A single Al KID deposited on a 2$\times$2\,cm$^2$ Si substrate. The chip was assembled in a copper structure using four PTFE elements.}
\end{figure}

The Aluminum prototypes, produced at Consiglio Nazionale delle Ricerche Istituto di Fotonica e Nanotecnologie (CNR-IFN, Rome, Italy), are fabricated using direct-write electron beam lithography (Vistec EBPG 5000). 
The resonators are deposited on high resistivity ($>$10\,k$\Omega\cdot$cm) Si(100) substrates. For the first detectors we used 275\,$\mu$m thick substrates, but we will investigate soon other thicknesses to better understand the efficiency in photons absorption and phonons propagation.
The Aluminum layer is evaporated in an electron-gun evaporator that allows to deposit the desired superconductor thickness (25 or 40\,nm for the detectors described later).
After the lift-off, we cut the Si wafer into a 2$\times$2\,cm$^2$ chip, that is assembled in a copper structure using insulating PTFE elements, as shown in Figure~\ref{fig:photo}.

The detectors are cooled below the critical temperature using a $^3$He/$^4$He dilution refrigerator with base temperature of about 10\,mK.
The output signal is fed into a CITLF4 SiGe low noise amplifier~\cite{amply}, which is thermally anchored to the 4\,K plate of the cryostat. The rest of the electronics is located at room temperature and its features, together with the description of the cryogenic facility and the acquisition software, can be found in references~\cite{CalderWhitePaper,Bourrion:2011gi,Bourrion:2013ifa}.

We developed several analysis tools to fit the complex transmission function S$_{21}$ in order to extract useful parameters, like the resonance frequency, the quality factor Q, the internal and coupling quality factors Q$_{int}$ and Q$_{c}$ (Q$^{-1}$=Q$_{int}^{-1}$+Q$_c^{-1}$).
Taking advantage from the analysis done by Swenson et al.~\cite{Swenson:2013jsa} and Khalil et al.~\cite{Khalil:2012}, we developed a model of the resonator that allows to extract the parameters of interest even in presence of large distortions produced, for example, by impedance mismatches between the feed-line and the chip~\cite{Casali:2015bhk}.

The resonance fit allows to determine also the center and radius of the resonance loop.
When an interactions occurs, the phase and amplitude of the resonance vary with respect to this reference frame, producing two readable signals. Since the phase signal $\delta\phi$ has a much better signal to noise ratio, we use this estimator to reconstruct the energy spectrum of the detector. 
According to our signal model, an energy release E produces a phase variation $\delta\phi$:
\begin{equation}
\delta\phi = \frac{\alpha S_2(\omega,T)}{N_0\Delta_0^2}\cdot\frac{Q}{V}\cdot\epsilon E\,.
\end{equation}
The first term in this formula is mainly material dependent: $\alpha$ is the fraction of kinetic inductance (of the order of 4--15$\%$ for Al), $S_2(\omega,T)$ is a slow function of the temperature (about 2.3--2.6 for our detectors), $N_0$ is the single spin density of states (for Al 1.72$\times$10$^{10}$\,eV$^{-1}\mu m^{-3}$), and 2$\Delta_0$ is the binding energy of Cooper pairs (about 200\,$\mu$eV for thin Al films). The second term, containing the resonator $Q$ and active volume $V$, depends mostly on the geometry of the device and can be easily varied up to an order of magnitude. Finally, $\epsilon$ is the detection efficiency.

\subsection{Efficiency optimization}
To measure the efficiency of our device, we performed several tests using a 2$\times$2\,cm$^2$, 275\,$\mu$m thick Silicon substrate. The substrate was illuminated with an optical source (400\,nm room-temperature LED coupled to an optical fiber) producing pulses up to $\sim$30\,keV, and with $^{55}$Fe/$^{57}$Co X-rays sources with peaks between 5.9 and 14.4\,keV, in order to cross-check the results. The calibration of the optical apparatus, indeed, is made with a photomultiplier at room temperature and is corrected using a Monte Carlo that accounts for the geometry of the set-up and for the optical properties of Si. On the contrary, the nominal energy of X-rays is known with high precision, allowing to investigate possible systematic errors in the calibration of optical pulses.
 All the sources were always placed on the back of the substrate to prevent direct illumination of the resonators.

Since we expect the efficiency to scale with the detector volume, we tested different KIDs geometries: (1) a 25\,nm thick Al KID with active area of 2.4\,mm$^2$,  (2) a 40\,nm thick KID with the same active area, and (3) a 40\,nm thick KID with an increased active area of 4.0\,mm$^2$. The results, reported in Table~\ref{tab:efficiency}, were obtained by illuminating the substrate in the proximity of the resonator, or as far as possible from it.

\begin{table}[thb]
\center
\caption{\label{tab:efficiency}Number of KIDs deposited on the 2$\times$2\,cm$^2$ Si substrate; thickness (t) and area (A) of the inductor; total detection efficiency measured with optical pulses illuminating the substrate region close to the KID ($\epsilon_{near}$) and far from the KID ($\epsilon_{far}$).}

\begin{tabular}{lccc}
KIDs          &t $\times$ A 	                &$\epsilon_{near}$    &$\epsilon_{far}$ \\					       							
number 	&[nm$\times$mm$^2$]       &[$\%$]                     &[$\%$] \\					              
\hline
1		&25$\times$2.4		&2			      &--\\
1		&40$\times$2.4		&7                           &6\\
1 		&40$\times$4.0		&11			      &8\\
\hline
4		&40$\times$2.4		&18                         &18\\
\end{tabular}
\end{table}

According to our predictions, the geometry with the largest active volume (geometry (3)) provides the highest detection efficiency.
The active volume of geometry (2) is about 60$\%$ of the one of geometry (3), and this results in a similar decrease of the efficiency.
On the contrary, given the ratio between the active volume of geometry (1) and geometry (3), we would have expected an efficiency of 4$\%$ for the 25\,nm KID.
Even if the errors on these values are rather large (about 20$\%$ for geometry (1) and 12$\%$ for geometry (2) and (3)), there is some tension between the expected number and the measured one, that has still to be understood. 
All these results were confirmed using the X-rays sources instead of the optical apparatus.

The results reported in Table~\ref{tab:efficiency} show also that the detection efficiency is lower when the source is placed far from the KID. This behavior can be corrected exploiting the time development of pulses: events occurring close to the KID show a larger amplitude, but also faster rise-time and faster-decay time.

Finally, we increased the number of resonators depositing 4 geometry (2) pixels on a single substrate. 
Summing the single efficiencies of 4 pixels of type (2) would result in an efficiency of 28$\%$, much larger with respect to the measured one (18$\%$).
This can be explained measuring the efficiencies of the single resonators: those placed in the nearby of the source show an efficiency similar to the one of the single pixel (6-7$\%$), while those far from the source feature a much lower efficiency, meaning that part of the signal was absorbed by the first pixels or lost through the substrate. 
Moving the source in another point of the substrate resulted in different efficiencies for the single pixels but, as shown in Table~\ref{tab:efficiency}, did not cause a significant variation of the \emph{total} efficiency of the detector.
This means that the 4-pixels configuration allows to efficiently collect the produced phonons.

As a final cross-check, we varied the resonator quality factor of one order of magnitude (from 10$^4$ to a few 10$^5$), we changed the coupling of the KIDs to the feed-line (from inductive to capacitive coupling) and we doubled the thickness of the resonators feed-line to investigate other possible systematics. The results were in full agreement with those reported in Table~\ref{tab:efficiency}, proving that our evaluation of the efficiency is robust.\\

\subsection{Noise optimization}
\begin{figure}[htbp]
 \includegraphics[width=.48\textwidth, natwidth=567, natheight=239]{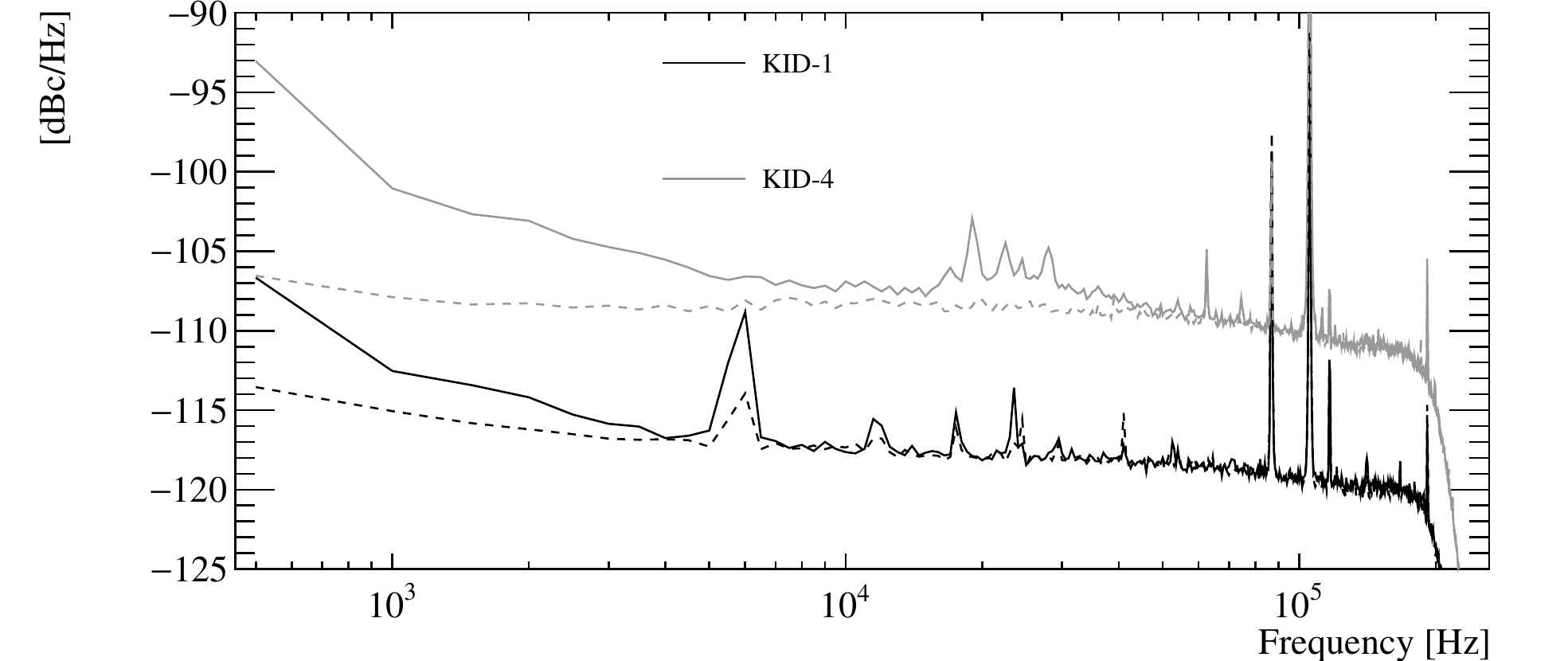}
 \caption{\label{fig:noise}Average noise power spectra for amplitude (dotted line) and phase (continuous line) read-out of a typical high-Q (KID-4) and a low-Q (KID-1) resonators.
 The flat noise is consistent with the amplifier noise. At low frequency another source of noise, not yet identified, appears in the phase noise power spectra, spoiling the energy resolution.}
\end{figure}

In parallel with the efficiency studies, we are performing a campaign devoted to the understanding and optimization of the detector noise.
In the ideal case, we would expect our noise power spectrum to be dominated by the cryogenic amplifier, which contribution to the resolution scales as:
\begin{equation}
\sigma_E \propto \frac{N_0\Delta_0^2}{\alpha S_2(\omega,T)}\frac{V}{Q\epsilon}\sqrt{\frac{kT_N}{P_f\tau_{qp}}}
\end{equation}
where $k$ is the Boltzmann constant, $T_N$ the noise temperature of the amplifier, P$_f$ the micro-wave power at the amplifier input and $\tau_{qp}$ the time constant in which quasi-particles recombine in Cooper pairs (of the order of hundreds of $\mu$s for these devices).
As explained in the previous section, some of these parameters, like the resonator Q or the active volume, can be varied up to one order of magnitude simply changing the pixel geometry or its coupling to the feed-line.
The optimization of the detector design allowed to improve the energy resolution from $\sim$150\,eV~\cite{Cardani:2015tqa} to about 90\,eV both in the single-pixel and in the 4-pixels configuration.
This value, which is already approaching the target of Phase--I (80 eV), could be further improved by suppressing the low-frequency noise that appears in all the prototypes that have been tested up to now (see Figure~\ref{fig:noise}).
This noise source is likely ascribable to the electronics and is now under investigation.

\section{More sensitive superconductors}
\label{sec:second phase}
In the next months we will continue the tests with Al-based prototypes to further optimize the detector design and to understand all the noise contributions.
In the meanwhile, we are starting the production of chips based on other, more sensitive superconductors.
The resolution, indeed, scales as:
\begin{equation}
\Delta E \propto \frac{T_C}{\epsilon \sqrt{QL}}
\end{equation}
thus, superconductors with lower critical temperature and higher inductance would allow to further enhance the sensitivity.
Possible candidates are sub-stoichiometric Titanium Nitride (TiN), or composite superconductors such as Ti+TiN or Ti+Al, which features are shown in Table~\ref{tab:superconductors}.
\begin{table}[thb]
\center
\caption{\label{tab:superconductors} Critical temperature T$_C$ and inductance $L$ for the superconductors of interest. The values of Al and TiN ($\alpha$ state) were measured using a film thickness of 40\,nm and 80\,nm respectively. The values of Ti+Al refer to a chip realized superimposing 10\,nm of Ti and 25\,nm of Al.}
\begin{tabular}{lcccc}
            		&Al   &TiN                           &Ti+TiN              &Ti+Al\\					       							
  	   		&      &sub-stoich.               &                        &\\					              
\hline
T$_C$		&1.2 &0.5				&0.5-0.8             &0.6-0.9\\
L [pH/square]   &0.5  &up to 50                  &6                      &1\\
\end{tabular}
\end{table}

We have already realized a single pixel detector with geometry (3) by depositing Ti+Al (10\,nm of Ti and 25\,nm of Al) on a 2$\times$2\,cm$^2$ Si substrate.
The first prototype, that was produced in collaboration with  CSNSM (Orsay, France) and Institut N\'eel, CNRS (Grenoble, France), is currently being tested and shows very encouraging results.

\section{Conclusion}
In this paper we presented a novel technology for the development of sensitive light detectors, based on phonon-mediated KIDs. 
We presented the results obtained depositing prototypes of Al KIDs with different geometries on 2$\times$2\,cm$^2$ Si substrates and we derived the detection efficiency. 
We evaluated the noise of the detectors and showed that the optimization of the design allowed to reach a baseline noise of about 90\,eV RMS. 
Finally, we discussed the perspectives of this technology, explaining how more sensitive superconductors will allow to reach the target sensitivity.

\section*{Acknowledgements}
This work was supported by  the European Research Council (FP7/2007-2013) under contract  CALDER no. 335359
and  by the Italian Ministry of Research under the  FIRB  contract no. RBFR1269SL.


\begin{thebibliography}{10}
\bibliographystyle{elsarticle-num} 

\expandafter\ifx\csname url\endcsname\relax
  \def\url#1{\texttt{#1}}\fi
\expandafter\ifx\csname urlprefix\endcsname\relax\def\urlprefix{URL }\fi
\expandafter\ifx\csname href\endcsname\relax
  \def\href#1#2{#2} \def\path#1{#1}\fi

\bibitem{Beeman:2013vda}
J.~W. Beeman, et~al., {Performances of a large mass ZnSe bolometer to search
  for rare events}, JINST 8 (2013) P05021.
\newblock \href {http://arxiv.org/abs/1303.4080} {\path{arXiv:1303.4080}},
  \href {http://dx.doi.org/10.1088/1748-0221/8/05/P05021}
  {\path{doi:10.1088/1748-0221/8/05/P05021}}.

\bibitem{Beeman:2012ci}
J.~W. Beeman, et~al., {Performances of a large mass ZnMoO(4) scintillating
  bolometer for a next generation 0vDBD experiment}, Eur. Phys. J. C 72 (2012)
  2142.
\newblock \href {http://arxiv.org/abs/1207.0433} {\path{arXiv:1207.0433}},
  \href {http://dx.doi.org/10.1140/epjc/s10052-012-2142-7}
  {\path{doi:10.1140/epjc/s10052-012-2142-7}}.

\bibitem{Cardani:2013dia}
L.~Cardani, et~al., {Development of a Li2MoO4 scintillating bolometer for low
  background physics}, JINST 8 (2013) P10002.
\newblock \href {http://arxiv.org/abs/1307.0134} {\path{arXiv:1307.0134}},
  \href {http://dx.doi.org/10.1088/1748-0221/8/10/P10002}
  {\path{doi:10.1088/1748-0221/8/10/P10002}}.

\bibitem{Beeman:2013zva}
J.~W. Beeman, et~al., {Characterization of bolometric Light Detectors for rare
  event searches}, JINST 8 (2013) P07021.
\newblock \href {http://arxiv.org/abs/1304.6289} {\path{arXiv:1304.6289}},
  \href {http://dx.doi.org/10.1088/1748-0221/8/07/P07021}
  {\path{doi:10.1088/1748-0221/8/07/P07021}}.

\bibitem{Bellini:2014yoa}
F.~Bellini, et~al., {Measurements and optimization of the light yield of a
  TeO$_2$ crystal}, JINST 9~(10) (2014) P10014.
\newblock \href {http://arxiv.org/abs/1406.0713} {\path{arXiv:1406.0713}},
  \href {http://dx.doi.org/10.1088/1748-0221/9/10/P10014}
  {\path{doi:10.1088/1748-0221/9/10/P10014}}.

\bibitem{Casali:2014vvt}
N.~Casali, et~al., {TeO$_2$ bolometers with Cherenkov signal tagging: towards
  next-generation neutrinoless double beta decay experiments}, Eur. Phys. J. C
  75~(1) (2015) 12.
\newblock \href {http://arxiv.org/abs/1403.5528} {\path{arXiv:1403.5528}},
  \href {http://dx.doi.org/10.1140/epjc/s10052-014-3225-4}
  {\path{doi:10.1140/epjc/s10052-014-3225-4}}.

\bibitem{Wang:2015taa}
G.~Wang, et~al., {R\&D towards CUPID (CUORE Upgrade with Particle
  IDentification)} (2015).
\newblock \href {http://arxiv.org/abs/1504.03612} {\path{arXiv:1504.03612}}.

\bibitem{Wang:2015raa}
G.~Wang, et~al., {CUPID: CUORE (Cryogenic Underground Observatory for Rare
  Events) Upgrade with Particle IDentification} (2015).
\newblock \href {http://arxiv.org/abs/1504.03599} {\path{arXiv:1504.03599}}.

\bibitem{Schaffner:2014caa}
K.~Schaeffner, et~al., {Particle discrimination in $TeO_2$ bolometers using
  light detectors read out by transition edge sensors}, Astropart. Phys. 69
  (2015) 30--36.
\newblock \href {http://arxiv.org/abs/1411.2562} {\path{arXiv:1411.2562}},
  \href {http://dx.doi.org/10.1016/j.astropartphys.2015.03.008}
  {\path{doi:10.1016/j.astropartphys.2015.03.008}}.

\bibitem{Willers:2014eoa}
M.~Willers, et~al., {Neganov-Luke amplified cryogenic light detectors for the
  background discrimination in neutrinoless double beta decay search with
  TeO$_{2}$ bolometers}, JINST 10~(03) (2015) P03003.
\newblock \href {http://arxiv.org/abs/1407.6516} {\path{arXiv:1407.6516}},
  \href {http://dx.doi.org/10.1088/1748-0221/10/03/P03003}
  {\path{doi:10.1088/1748-0221/10/03/P03003}}.

\bibitem{Biassoni:2015eij}
M.~Biassoni, et~al., {Large area Si low-temperature light detectors with
  Neganov Luke effect}, Eur. Phys. J. C 75~(10) (2015) 480.
\newblock \href {http://arxiv.org/abs/1507.08787} {\path{arXiv:1507.08787}},
  \href {http://dx.doi.org/10.1140/epjc/s10052-015-3712-2}
  {\path{doi:10.1140/epjc/s10052-015-3712-2}}.

\bibitem{CalderWhitePaper}
E.~Battistelli, F.~Bellini, C.~Bucci, M.~Calvo, L.~Cardani, N.~Casali, M.~G.
  Castellano, I.~Colantoni, A.~Coppolecchia, C.~Cosmelli, et~al., {CALDER:
  Neutrinoless double-beta decay identification in TeO$_2$ bolometers with
  kinetic inductance detectors}, Eur. Phys. J. C 75~(8).
\newblock \href {http://arxiv.org/abs/1505.01318} {\path{arXiv:1505.01318}}.

\bibitem{Day:2003fk}
P.~K. Day, H.~G. LeDuc, B.~A. Mazin, A.~Vayonakis, J.~Zmuidzinas, A broadband
  superconducting detector suitable for use in large arrays, Nature 425~(6960)
  (2003) 817--821.
\newblock \href {http://dx.doi.org/10.1038/nature02037}
  {\path{doi:10.1038/nature02037}}.

\bibitem{swenson}
L.~J. Swenson, A.~Cruciani, A.~Benoit, M.~Roesch, C.~S. Yung, A.~Bideaud,
  A.~Monfardini, {High-speed phonon imaging using frequency-multiplexed kinetic
  inductance detectors}, Appl.Phys.Lett. 96 (2010) 263511.
\newblock \href {http://arxiv.org/abs/1004.5066} {\path{arXiv:1004.5066}},
  \href {http://dx.doi.org/10.1063/1.3459142} {\path{doi:10.1063/1.3459142}}.

\bibitem{moore2}
D.~C. Moore, S.~R. Golwala, B.~Bumble, B.~Cornell, P.~K. Day, H.~G. LeDuc,
  J.~Zmuidzinas, {Position and energy-resolved particle detection using
  phonon-mediated microwave kinetic inductance detectors}, Appl.Phys.Lett. 100
  (2012) 232601.
\newblock \href {http://arxiv.org/abs/1203.4549} {\path{arXiv:1203.4549}},
  \href {http://dx.doi.org/10.1063/1.4726279} {\path{doi:10.1063/1.4726279}}.

\bibitem{amply}
\href{http://radiometer.caltech.edu/datasheets/amplifiers/CITLF4.pdf}{{CITLF4
  SiGe Data Sheet}}.
  \newline http://radiometer.caltech.edu/datasheets/amplifiers/CITLF4.pdf

\bibitem{Bourrion:2011gi}
O.~Bourrion, A.~Bideaud, A.~Benoit, A.~Cruciani, J.~F. Macias-Perez,
  A.~Monfardini, M.~Roesch, L.~Swenson, C.~Vescovi, {Electronics and data
  acquisition demonstrator for a kinetic inductance camera}, JINST 6 (2011)
  P06012.
\newblock \href {http://arxiv.org/abs/1102.1314} {\path{arXiv:1102.1314}},
  \href {http://dx.doi.org/10.1088/1748-0221/6/06/P06012}
  {\path{doi:10.1088/1748-0221/6/06/P06012}}.

\bibitem{Bourrion:2013ifa}
O.~Bourrion, C.~Vescovi, A.~Catalano, M.~Calvo, A.~D'Addabbo, J.~Goupy,
  N.~Boudou, J.~F. Macias-Perez, A.~Monfardini, {High speed readout electronics
  development for frequency-multiplexed kinetic inductance detector design
  optimization}, JINST 8 (2013) C12006.
\newblock \href {http://arxiv.org/abs/1310.5891} {\path{arXiv:1310.5891}},
  \href {http://dx.doi.org/10.1088/1748-0221/8/12/C12006}
  {\path{doi:10.1088/1748-0221/8/12/C12006}}.

\bibitem{Swenson:2013jsa}
L.~J. Swenson, P.~K. Day, B.~H. Eom, H.~G. Leduc, N.~Llombart, C.~M. McKenney,
  O.~Noroozian, J.~Zmuidzinas, {Operation of a titanium nitride superconducting
  microresonator detector in the nonlinear regime}, J. Appl. Phys. 113 (2013)
  104501.
\newblock \href {http://arxiv.org/abs/1305.4281} {\path{arXiv:1305.4281}},
  \href {http://dx.doi.org/10.1063/1.4794808} {\path{doi:10.1063/1.4794808}}.

\bibitem{Khalil:2012}
M.~S. Khalil, M.~J.~A. Stoutimore, F.~C. Wellstood, K.~D. Osborn, {An analysis
  method for asymmetric resonator transmission applied to superconducting
  devices}, J. Appl. Phys. 111 (2012) 054510.
\newblock \href {http://dx.doi.org/10.1063/1.3692073}
  {\path{doi:10.1063/1.3692073}}.

\bibitem{Casali:2015bhk}
N.~Casali, et~al., {Characterization of the KID-Based Light Detectors of
  CALDER}, Journal of Low Temp. Phys.\href {http://arxiv.org/abs/1511.05038}
  {\path{arXiv:1511.05038}}, \href
  {http://dx.doi.org/10.1007/s10909-015-1358-y}
  {\path{doi:10.1007/s10909-015-1358-y}}.

\bibitem{Cardani:2015tqa}
L.~Cardani, et~al., {Energy resolution and efficiency of phonon-mediated
  Kinetic Inductance Detectors for light detection}, Appl. Phys. Lett. 107
  (2015) 093508.
\newblock \href {http://arxiv.org/abs/1505.04666} {\path{arXiv:1505.04666}},
  \href {http://dx.doi.org/10.1063/1.4929977} {\path{doi:10.1063/1.4929977}}.

\end{thebibliography}
\end{document}